# Antenna parameterization for effectiveness in horn shaped antenna for 5G communication as future of Antennas

Sunit Shantanu Digamber Fulari,

**ECE Department, Chandigarh University**

**Abstract:** Horn antenna is well documented in our research in this paper. We are trying our latent method of radiation by antennas which we suppose to reduce to a significant extent. Why we chose horn antenna is it resonates the sound to explosion as done by horn shaped matter. Horn by its shape makes the sent signals to maximum capability by its shape which is required by the receiver due to the distance separation from sender and receiver. Our research will contain various implementations leading to improvement of previous designs. We will use the traditional methods of Bergen to make the antenna behave smartly in its functioning.

**I. Introduction:**

Horn shaped is a kind of antenna which blows the signals at its base of expulsion of waves energy. This energy gets sent from the antenna, due to its shape it blows out energy into the atmosphere to the receiver.

We demonstrated our analysis of radiation pattern for our horn antenna for the following set of frequency. More the frequency more is the radiation of the antenna.

At a 2.4Ghz frequency we have

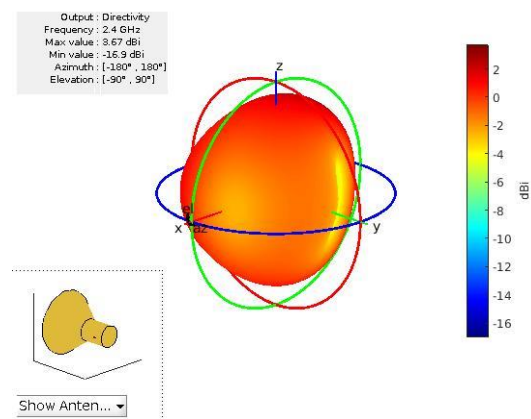

Figure I:

This is a starting latent frequency which can be considered to have reverse frequency of the radiation, radiation is shown going in center direction to progression.

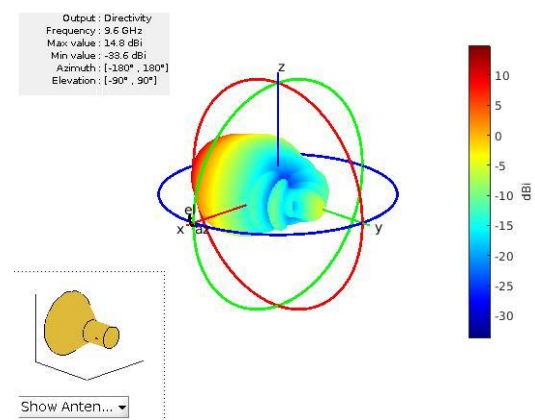

Figure II: Looking like funnel and corrugated shape.

9.6 Ghz as above shown a unique characteristics which can be considered to be having side lobes and main lobes [1][5]. The green is the side lobe line while red is the back lobe. This is a peculiar design which shows improvement of design as radiation of the antenna is shown to migrate to minimum. This shows a tendency of the antenna to function with maximum power in the forward direction and cause the shape of the antenna is horn shaped it propels the signals in exorbitant force compared to linear antenna. This seems to be a bit confusing but further analysis will solve the problem[6][8].

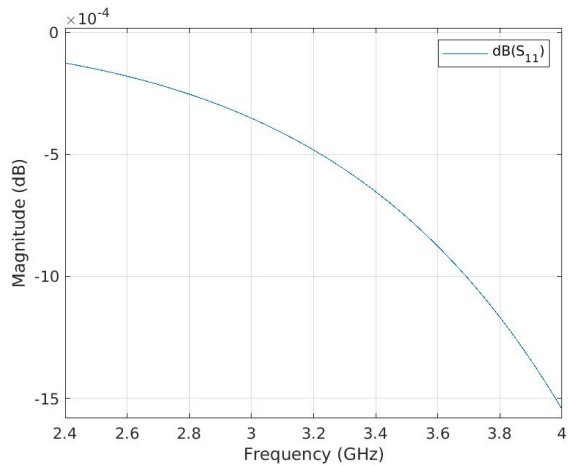

Figure III:

The S parameter is shown to decrease non uniformly but exponentially with increase of frequency. Impedance is as above which shows the resistance of the entire circuit of our design model [9][15].These antennas are primary used in short distances where the intensity to be received is large. They can be used in military operations where there are single digit miles to be covered in distance, this is in contrary to linear antennas. Horn antenna is smooth in the beginning but expands at the end where the energy has to be sent and expanded to be sent to the receiver. 5G is a upgraded version of 4G with impeded and advanced radiation, that is the antennas are made to operate at advanced frequency with new set of antennas. Our horn shaped antenna will be used in 5G communication which deals with speeded communication and interstellar space communication as they are also as aperture antennas. There is no time lag, or the time lag is in microsecond, this will enable exchanging of communication between far away countries without any time lag. Horn antennas were studied during the last century but they lost thorough study due to the development of variousu other versatile antennas and systems. This Horn shaped antenna can be proved very fruitful in the long run due to its shape which forms simplicity at the same time usefulness in functioning. Indeed the corrugated shape and expanded shape both prove to make this as a very good antenna function, which can be used in military, industry, short distance communication varying in importance, commercial , and in more applications in which the shape itself proves to aid in communication. This figure V is a typical horn antenna used for study. It proves futile if there is no expansion during transmission of energy in the case of pipe based or rod based design. Next mutant in this variety is a rod shaped antenna with a expansion at the other end unlike the funnel or corrugated shaped antenna. This also proves to be quite effective in short distance communication.

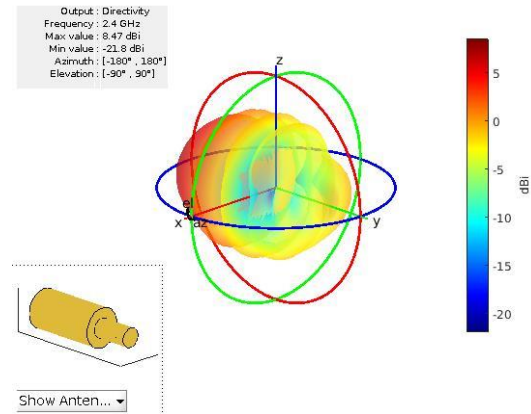

Figure V:

The 2.4Ghz modelled horn antenna known with a more common name as tappered, pyramidal and more frequenctly as potter.

Horn antenna was a not so frequent group of study until it was published by Wang et.al in his paper spoke about major things in his paper about dielectric loaded substrate. In this paper they have very interestingly spoken things intelligently about finer elements in fabrication and simulation of horn antennas.

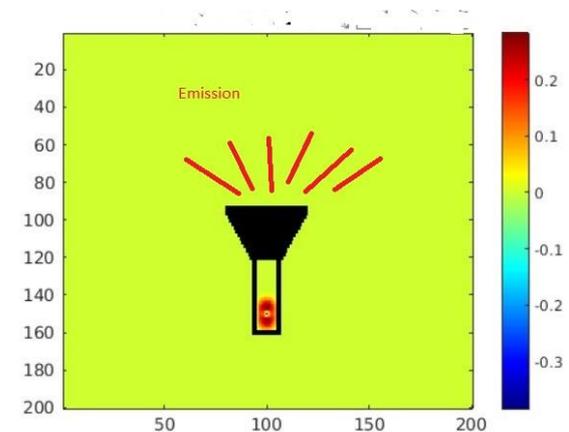

Figure VI:Horn shaped antenna for simulation.

The waves start from the bottom of the antenna just to be expelled at the mouth of the antenna. Few like Morote et.al in his paper have spoken about H plane radiation into consideration in horn shaped antenna. Scrimping is a phenomena of compressing the mouth to helical form to forn a antenna. We want this antenna to behave smartly and radiate only when there is a motion seen in the vicinity of the antenna, in this way there will be reduced radiation energy from the antenna making the antenna bahave smartly reducing radiation in the area for its operation. We are going to use mathematical models in this estimation which detects changes in motion of the object for infinitesimal changes in any body. Its clear that we will have to use a camera with high pixels and depth to detect objects in its vicinity secondly when the object is not in range we have come up with a algotihm derived by bergen at.al which speaks about optical flow and object detection by differential equation. This is complex in its derivation in which we try to derive a whole set of equations in our proposal. Figure I shows a radiation pattern which is concentrated at the centre all the lobes while figure II shows us a comparison of back and side lobe to be more efficient and proficient at the same time whilst the figure V determines a unique parameters of back lobe which handles the radiation of energy due to back lobe shown in red. Our functioning of the antenna is without a camera, this is the interesting part of our analysis. The antenna will check the presence of devices on the basis of its energy which is getting used up or absorbed, accepted based on the presence of elements in the vicinity of the antenna.

**Proposed Theory:**

This is a unique basis of antenna functioning which determines radiation transmission only when there is a client in the 6-8 miles of the antenna feed element of the horn antenna. Today we know this radiation is used to find and broadcast the video on the display device just on the basis of radiation by the antenna. The devices have started to implement based on the latest research. Well, our proposal in this paper in on Horn antenna which will have a control in a certain area absorbing the area by its radiation and when there is a certain object in its vicinity it will send more radiation in that direction. This concept is similar to the antenna being compared to a car with a driver inside it, the entire device which is the car is under the control of the driver when he is operating it similarly we want the antenna to be under the control of the base antenna or the controlling device which will control the radiation, and this radiation will be the car keeping the antenna as the driver in comparison to a driver in a car. Our novel idea will use the horn type antenna in land communication, we already call this antenna as having an aperture kind of design. Indeed we are very close in mart radiation of the antenna, it will have a particular range in its control. That is the example is similar to a broadband modem which operates in a house, restaurant, or offices, etc when it has the whole setup under its control and forms a network for data exchange, as this is the basic system which forms the basis of operation of the modern system of communication using the antennas. With modern developments of visual radiation we can find out a region of space by monitoring a place later viewing it on the television due to the radiation given out by the antenna of the modem. Moallemizadeh et.al in this paper has implemented reduction of side lobe based on the technique by using horn antenna is their famous paper names "A simple design substrate-integrated waveguide horn antenna with reduced back lobe" this is a good paper in horn antennas which shows a good design performance compactness at 18.7 *23.34*7.62 mm(0.93*1.16*0.37 $\lambda_0$ ) and the measured gain of the horn is about 5.5 dB at 15GHz at a good bandwidth of six percent. Smartness of the antenna is on the way it operates on radiation.

**Summary of our concept:**

The radiation and feed with the design forms the entire system of our study. This radiation energy is to be smartly controlled by the antenna by feeling smartly who is absorbing, that is coming under its influence. So this smart functioning of the antenna is to be implemented in our paper.

We tried to simulate the horn antenna at 1.645 frequency.The results obtained were as below.

The simulation shown as below showed that there was uniform decrease in far field with the increase of frequency. This marked a significant understanding that the antenna will perform better with the increase of frequency as the gain decreased. This gain understood to show how much the power is in radiation of the antenna in that direction. We want maximum radiation in the defined direction when it is required for any purposes, that is when there is connection between the receiver and the transmitter or sender. We are trying to make the antenna behave smartly with our problem. We have used cadfeko in our simulation of horn antenna to obtain polar graphs and far field regions for the functioning of the antenna in various circumstances. Considering plain and hilly areas of operation of the omnidirectional antenna we come to the analysis that this antenna with high gain will be better in operation.

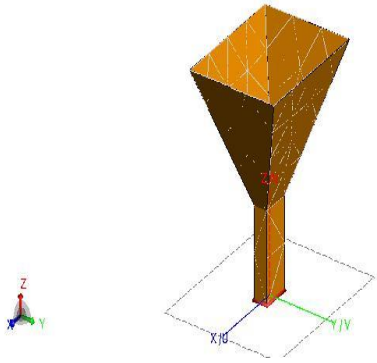

Figure VI:simulated antenna horn using CAD Feko.

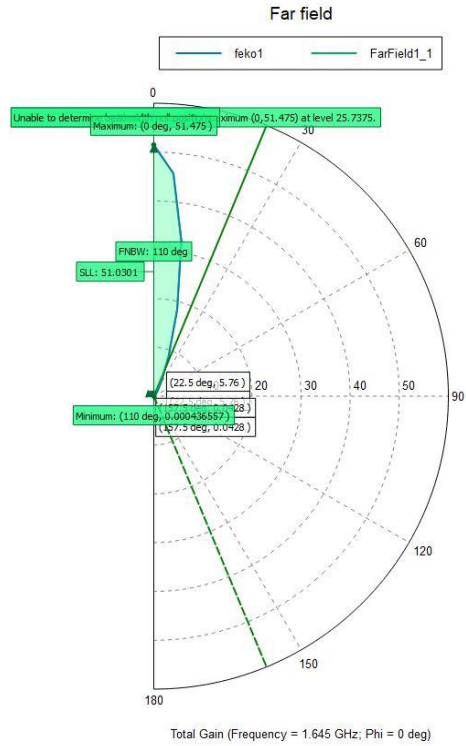

This figure above VIII makes it clear that far field plot which shows the magnitude of radiation in defined direction, this states that there will be increased radiation in given direction.

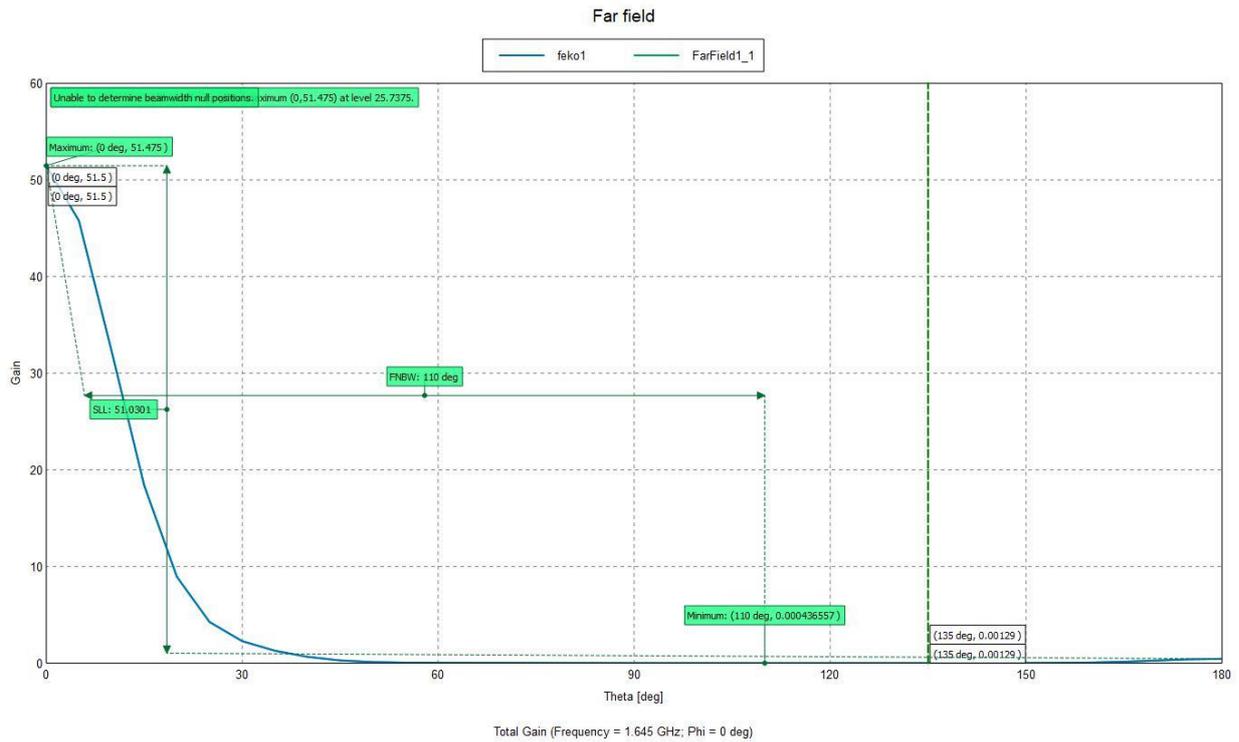

Figure IX: This shows that gain decreases with increase in frequency, that is the power of radiation also decreases when it is not required to be performing the task.

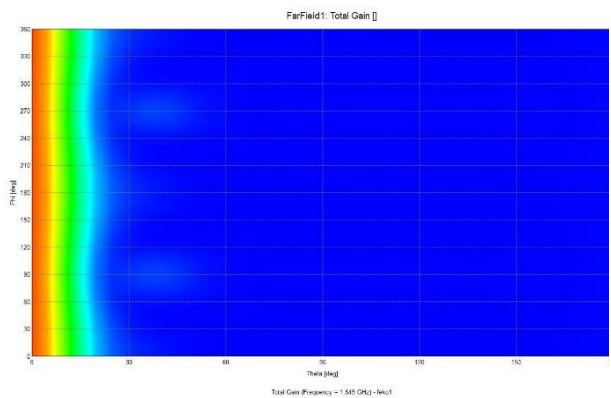

Figure X: Shows near field region for a frequency of 1.545. The presence of blue light inself shows that this antenna will be very efficient due to the frequency.

**Conclusion:** We have simulated the horn shaped antenna for the operation in different frequencies by studying the far field, near field, cartesian and polar plots. We have also tried to put forth out proposition of smart antennas, which creates a wide scope for further study in this topic of area. We have also tried to study radiation pattern of this horn antenna for various energy of operation. We have simulated the radiation pattern for varying frequency from 2.4GHz to 9.6 GHz. There is lot of scope for further study in this type of aperture antenna which creates lot of interest for further researchers in this field.